\documentclass[sigconf]{acmart}
\hypersetup{draft}

\usepackage[T1]{fontenc}
\usepackage{afterpage}
\usepackage{listings}
\usepackage{xcolor}
\usepackage{graphicx}
\graphicspath{{figs/}{figs/model/}{figs/psucc/}}
\usepackage{float}
\usepackage[font=small]{caption}
\usepackage[font=footnotesize]{subcaption}
\usepackage{placeins}
\usepackage{url}
\usepackage{epstopdf}
\usepackage{booktabs}
\usepackage{footnote}
\makesavenoteenv{table}
\makesavenoteenv{tabular}
\usepackage{tikz}
\usepackage{pgfplots}
\usepackage{dblfloatfix} 

\usepackage[acronyms,nonumberlist,nopostdot,nomain,nogroupskip]{glossaries}

\usepackage{tabularx}

\usepackage{paralist}

\newacronym{phy}{PHY}{Physical}
\newacronym{mac}{MAC}{Medium Access Control}
\newacronym{ns}{NS}{Network Server}
\newacronym{gw}{GW}{Gateway}
\newacronym{ed}{ED}{End Device}
\newacronym{adr}{ADR}{Adaptive Data Rate}
\newacronym{sf}{SF}{Spreading Factor}
\newacronym{ack}{ACK}{Acknowledgment}
\newacronym{iot}{IoT}{Internet of Things}
\newacronym[plural=LPWANs,firstplural=Low Power Wide Area Networks (LPWANs)]{lpwan}{LPWAN}{Low Power Wide Area Network}
\newacronym{ul}{UL}{uplink}
\newacronym{dl}{DL}{downlink}
\newacronym{qos}{QoS}{Quality of Service}
\newacronym{css}{CSS}{Chirp Spread Spectrum}
\newacronym{dc}{DC}{Duty Cycle}
\newacronym{rx1}{RX1}{first receive window}
\newacronym{rx2}{RX2}{second receive window}
\newacronym{fdgw}{FDGW}{Full Duplex Gateway}
\newacronym{ttn}{TTN}{The Things Network}
\newacronym{ism}{ISM}{Industrial, Scientific, and Medical}
\newacronym{lora}{LoRa}{Long-Range}
\newacronym{toa}{ToA}{time on air}
\newacronym{cpsr}{CPSR}{Confirmed Packet Success Rate}
\newacronym{uu}{UU}{Unconfirmed Uplink PDR}
\newacronym{cu}{CU}{Confirmed Uplink PDR}
\newacronym{cd}{CD}{Confirmed Downlink PDR}
\newacronym{pdr}{PDR}{Packet Delivery Rate}
\newacronym{per}{PER}{Packet Error Rate}
\newacronym{ber}{BER}{Bit Error Rate}
\newacronym{mcu}{MCU}{Micro Controller Unit}
\newacronym{dr}{DR}{Data Rate}

\begin{document}


\flushbottom
\setlength{\parskip}{0ex plus0.1ex}
\addtolength{\skip\footins}{-0.2pc plus 40pt}

\title{An ns-3 implementation of a battery-less node \\ 
for energy-harvesting Internet of Things}

\author{Martina Capuzzo}
\email{capuzzom@dei.unipd.it}
\affiliation{%
  \institution{Department of Information Engineering, \\ University of Padova}
  \city{Padova}
  \country{Italy}
}
\author{Carmen Delgado}
\email{carmen.delgado@i2cat.net}
\affiliation{%
  \institution{i2CAT Foundation}
  \city{Barcelona}
  \country{Spain}
}

\author{Jeroen Famaey}
\email{jeroen.famaey@uantwerpen.be}
\affiliation{%
  \institution{IDLab, \\ University of Antwerp - imec}
  \country{Belgium}
}

\author{Andrea Zanella}
\email{zanella@dei.unipd.it}
\affiliation{%
  \institution{Department of Information Engineering, \\ University of Padova}
  \city{Padova}
  \country{Italy}
}

\copyrightyear{2021}
\acmYear{2021}
\setcopyright{acmlicensed}
\acmConference[WNS3 2021]{2021 Workshop on ns-3}{June 23--24, 2021}{Virtual conference}
\acmBooktitle{2021 Workshop on ns-3 (WNS3 2021), June 23--24, 2021, Virtual conference}
\acmPrice{15.00}
\acmDOI{xxxxxx.xxxxxxxxxxxxxx}
\acmISBN{xxxxxxxxxxxxxxxxxxxxxxxx}

\pagestyle{empty}

\begin{abstract}
In the \gls{iot}, thousands of devices can be deployed to acquire data from the environment and provide service to several applications in different fields. In many cases, it is desirable that devices are self-sustainable in terms of energy.  
Therefore, the research community is exploring the possibility of employing battery-less devices, where the energy is derived solely from external and/or environmental sources, such as solar panels. In this work, we propose an ns-3 model of a (super) capacitor, which can be used as the storage of the harvested energy in a battery-less \gls{iot} device, and add the support for the intermittent behavior of devices, turning off/on according to their energy level. To exemplify the use of the model, we apply it to a LoRaWAN node, and compare the simulation outcomes with results in the literature obtained with mathematical analysis, confirming the accuracy of the implementation. Then, we show the importance of analyzing the interaction between energy availability and communication performance, paving the way for more accurate and realistic simulations in the field.
The implemented code is made available as open source.

\end{abstract}

\ccsdesc[500]{Networks~Network simulations}
\ccsdesc[500]{Sensor Networks}

\keywords{Network simulations, ns-3, Internet of Things, Battery-less device, Capacitor, Energy Harvesting} 

\maketitle

\section{Introduction}
\label{sec:intro}
\glsresetall
In the last years, the \gls{iot} paradigm has been applied in many contexts, such as Smart Cities, healthcare, and industrial and agricultural scenarios. In such application scenarios, the network infrastructure must be able to support a large number of devices, generally transmitting at low bit rates. The communication technologies, in turn, need to serve wide networks with a large number of devices. Therefore, in the last years, \gls{lpwan} technologies have been proposed, among which LoRaWAN and Sigfox have reached a large popularity. 

In many cases, devices are battery-powered and, although the communication technology is designed to limit the energy consumption, adverse network conditions, channel impairments or unwise settings can negatively affect the device's lifetime. Unfortunately, battery replacing is costly from an economic and environmental perspective, pushing for a migration to green solutions. One possibility is to use energy harvesting techniques, which derive energy from renewable sources (e.g., solar power, thermal/wind energy), and can stored it in (super) capacitors integrated in battery-less devices. Nevertheless, the inconstant behavior of the harvested energy impacts on the device's capabilities, including communication. 

This aspects have started gaining interest in the last few years, but have not been deeply evaluated yet. 
Therefore, works that deal with \gls{iot} networks with battery-less devices and energy harvesting~\cite{fraternali2018pible, delgado2020battery, sabovic2020energy, gindullina2020energy}, mainly address theoretical analysis (i.e., mathematical modeling) or empirical evaluations. Instead, the use of simulations can provide a precious help in the evaluation of the interplay between the network state, the system configuration and the device's energy capabilities. For example, the communication could benefit from the robustness provided by message repetitions or the use of a higher transmission power, but this will impact the energy autonomy of the device; conversely, low energy levels may prevent the correct transmission of some packets. These aspects are further complicated when considering the variability of the energy source in nodes with harvested power and the interference of many communicating devices.

In this paper, we propose an ns-3 implementation of a battery-less node with a (super) capacitor coupled with an energy harvester, which enables the performance evaluation of \gls{iot} networks with battery-less devices. This is built as an extension of the native ns-3 energy model, and can be easily integrated with existing ns-3 modules, supporting the intermittent behavior of devices that can turn off or on according to their energy level. Also, a supplementary class makes it possible to import values for the harvested power from external files, which can be obtained from real measurements. The implemented code is available at~\cite{ns3capacitor}. 

We validate the proposed framework by leveraging the \texttt{lorawan} ns-3 module, and describe some results obtained when considering a LoRaWAN battery-less device.

The reminder of this paper is structured as follows. Section~\ref{sec:model} describes how the battery-less device is modeled, while the specific ns-3 implementation and integration with existing modules are discussed in Section~\ref{sec:implementation}. In Section~\ref{sec:validation} we first briefly describe the LoRaWAN technology that we use in our simulations to validate the capacitor's implementation, and then discuss the obtained results. Finally, Section~\ref{sec:conclusions} reports the conclusions and possible extensions of this work.

\section{Batteryless IoT devices with Energy Harvesting}
\label{sec:model}
\begin{figure}[t]
\centering
    \begin{picture}(100,120)
    \put(-50,20){\includegraphics[width=0.8\linewidth]{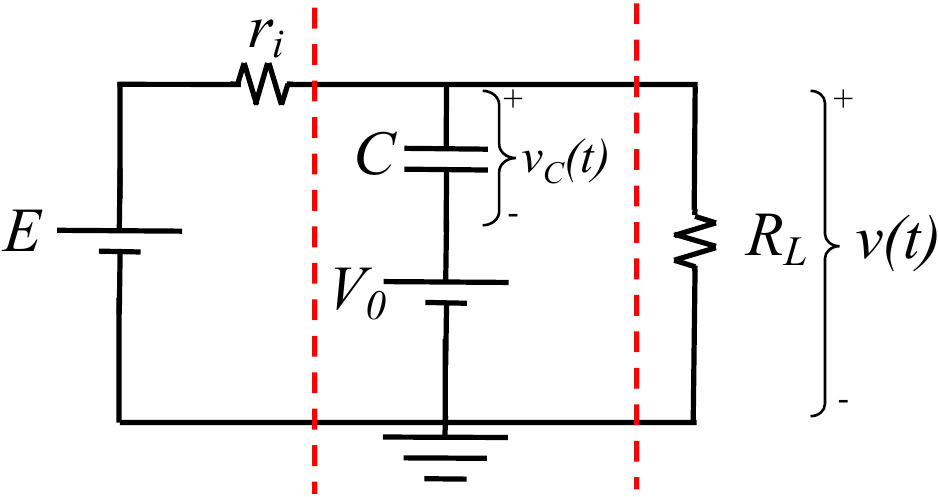}}
    \put(-30,10){Harvester}
    \put(30,10){Capacitor}
    \put(90,10){Load: CPU, Radio,...}
    \end{picture}
    \caption{Electrical circuit model of a battery-less IoT device~\cite{delgado2020battery}.}
    \label{fig:circuit}
\end{figure}
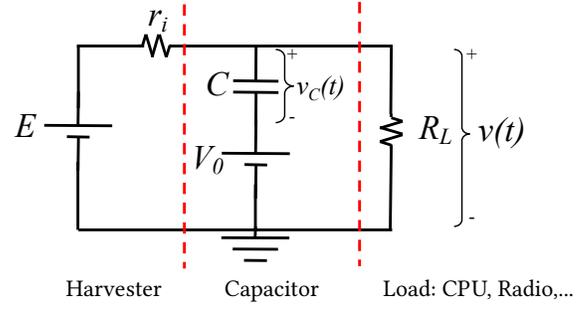

To model a battery-less \gls{iot} device, we consider approach used in~\cite{delgado2020battery}. The device consists of several components: a \gls{mcu}, a radio unit, some peripherals (e.g., for sensing purposes), a capacitor to store energy, and a harvester mechanism to recharge the capacitor. The overall device can be modeled as an equivalent electrical circuit with three parts: (i) the harvester, (ii) the capacitor, and (iii) the load, as represented in Fig.~\ref{fig:circuit}, and better described next.
%

\begin{figure}[t]
  \centering
    \includegraphics[width=0.43\textwidth]{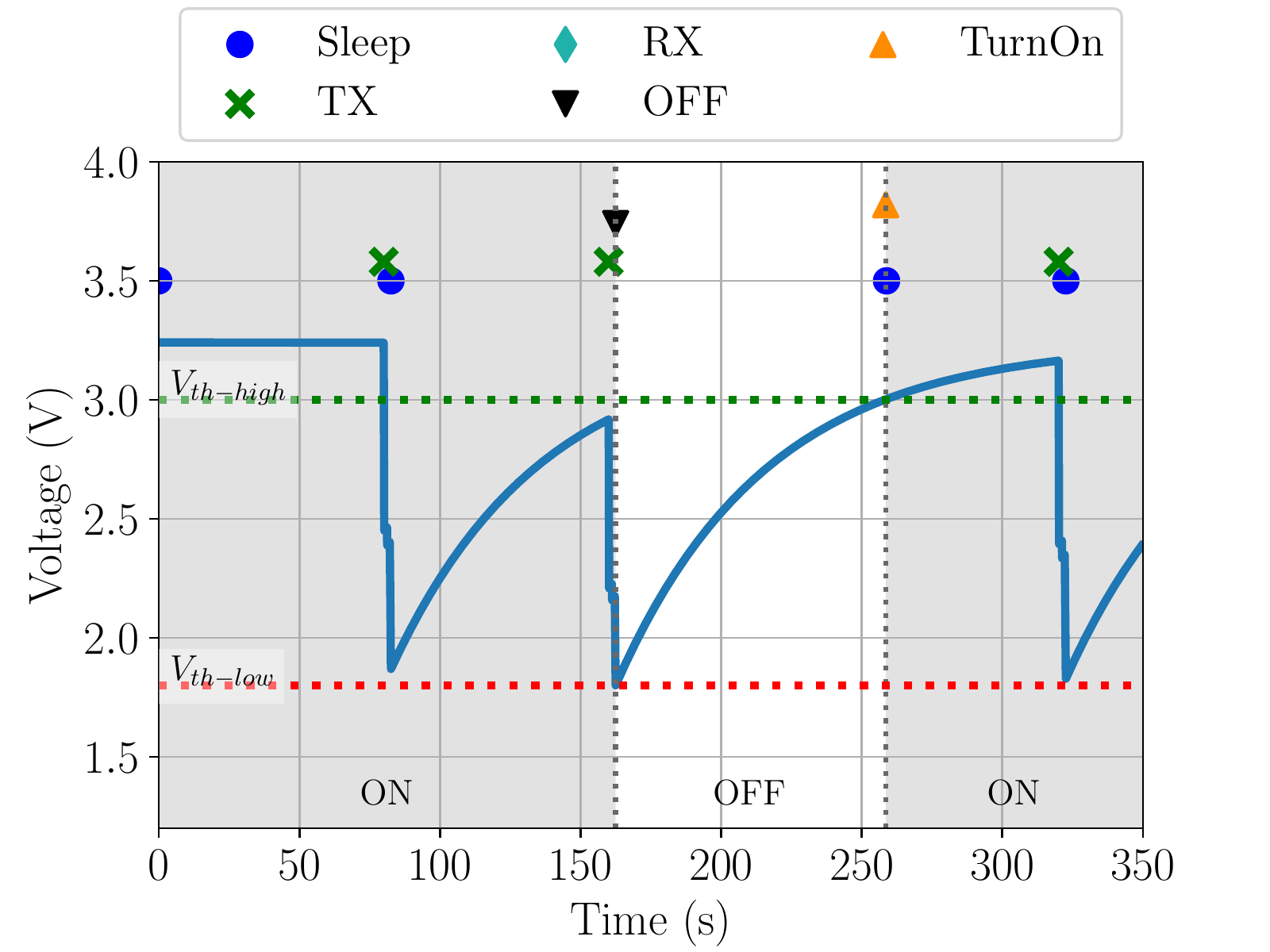}
  \caption{Example of the device's voltage when it enters different phases, with $P_{harvester}=0.001$~W, packet generation period of 60~s, and LoRa DR~3.}
  \label{fig:timevoltage}
\end{figure}

\begin{enumerate}[(i)]
\item \emph{The harvester:} it is the only energy source in the system. 
The harvester is modelled as an ideal constant voltage source (denoted by $E$) with a series resistance ($r_i(t)$) that determines the maximum power that can be produced by the harvester, which is given by
\begin{equation}
P_{harvester}(t) = \frac{E^2}{r_i(t)}\,.
\end{equation}
In general, the resistance $r_i(t)$ can change in time, according to the fluctuations on the energy harvesting process. By coupling the harvester with a voltage regulator, however, the output voltage $E$ can be stabilized. Our model makes it possible to either generate the harvested power values as independent random samples taken from a given distribution, or to read them from a pre-loaded trace file. 
%
\item \emph{The load:} it models all the components of the system that consume energy. According to the activity performed by each component, it is possible to define different \emph{states} of the load, which are characterized by a specific power consumption. For each state, we can therefore define a load resistance $R_L(s)$, which is computed considering the total current $I_{load}(s)$ absorbed by the load in the specific state $s$:
\begin{equation}
    R_L(s) = \frac{E}{I_{load}(s)}.
\end{equation}
\item \emph{The capacitor:} it stores the energy generated by the harvester and releases it to the load when required. The behavior of the system can be represented by a series of intervals corresponding to different events/activities (e.g., \gls{mcu} active and radio transmitting), corresponding to the different states of the load. For each state $s$, the voltage of the capacitor can be represented by $(V_0, v_C(t))$, where $V_0$ is the capacitor voltage when entering the state, and $v_C(t)$ is the voltage of the capacitor after $t$ seconds spent in state $s$. As depicted in Fig.~\ref{fig:circuit}, $V_0$ is included in the circuit as an ideal voltage source, while $v_C(t)$ is the voltage over time of an ideal capacitor. 

The voltage provided by the capacitor to the load after $t$ seconds in state $s$ can thus be computed as 
\begin{equation}
    v(t, s)= E \frac{R_{eq}(s, t)}{r_i(t)}\big(1 - e^{-\frac{t}{R_{eq}(s, t)C}}\big) + V_0 e^{-\frac{t}{R_{eq}(s, t)C}},
    \label{eq:vc}
\end{equation}
where $C$ is the capacitance of the capacitor [in Farads], and 
\begin{equation}
    R_{eq}(s, t) = \frac{R_L(s)r_i(t)}{R_L(s) + r_i(t)}.
\end{equation}

\end{enumerate}
To model real devices, we consider that they may switch off at anytime because of an energy level too low to continue their functioning. Therefore, we define two voltage thresholds for $v(t, s)$: $V_{th\_low}$, below which the device switches off, and $V_{th\_high}$, above which the device goes back to active state, thanks to the harvested energy. As an example, Fig.~\ref{fig:timevoltage} shows the voltage level of a device over time, highlighting the on/off phases with $V_{th\_low} = 1.8~V$ and $V_{th\_high} = 3~V$. The specific states the device goes through will be better explained in Section~\ref{sec:validation}.

\section{Code implementation in ns-3}
\label{sec:implementation}
\subsection{Capacitor}
The ns-3 class implementing the storage of the energy in a capacitor is called \texttt{CapacitorEnergySource}, as it extends the \texttt{EnergySource} class available in ns-3.\footnote{\url{https://www.nsnam.org/doxygen/classns3_1_1_energy_source.html}} Thus, it is used as done for the classes implementing the Lithium Ion  Battery or the non-linear battery model. As such, it can be easily connected to energy harvester components (\texttt{EnergyHarvester}) and to the class modeling the energy consumption behavior of the device (\texttt{DeviceEnergyModel}).\footnote{To avoid compilation errors, in \texttt{ns3/src/energy/model/energy-source.h}, the \emph{private} variables should be moved to \emph{protected}.} The capacitor's features can be set using the class attributes and methods; the most relevant attributes are reported in~Table~\ref{tab:attributes}.
\begin{table*}[]
    \centering
    \begin{tabular}{p{0.3\textwidth}p{0.6\textwidth}}
    \toprule
        Attribute  &  Description \\
    \midrule
        Capacitance & Capacitance [F] \\
        CapacitorEnergySourceInitialVoltage & Initial voltage of the capacitor [V] \\
        CapacitorMaxSupplyVoltage & Maximum supply voltage for the capacitor energy source [V] \\
        CapacitorLowVoltageThreshold & $V_{th\_low}$, as fraction of the maximum supply voltage\\
        CapacitorHighVoltageThreshold & $V_{th\_high}$, as fraction of the maximum supply voltage \\
        PeriodicVoltageUpdateInterval & Time interval between periodic voltage updates \\
        \bottomrule
    \end{tabular}
    \caption{Relevant attributes of the \texttt{CapacitorEnergySource} class.}
    \label{tab:attributes}
\end{table*}

The value of the voltage stored in the capacitor can be updated periodically by calling the appropriate function: \texttt{UpdateEnergySource}. This function computes $v(s, t)$ as for Eq.~(\ref{eq:vc}) according to the current state $s$, with $V_0$ being the voltage computed at its previous call. It is recommended to call the function also before switching the device to a new state, in order to keep the voltage up-to-date with the correct value of $R_L(s)$ and guarantee that the device's energy is not depleted, preventing the correct switching to the new state.

Besides the common ``setters'' and ``getters'' methods, and the auxiliary functions to compute the voltage level, the class also provides the following functions:
\begin{itemize}
    \item \texttt{IsDepleted}, returning \texttt{true} if the current voltage value is below $V_{th\_low}$;
    \item \texttt{ComputeLoadEnergyConsumption}, which computes the energy dissipated only by the load where a given current flows for a given time interval, and the initial voltage level of the capacitor is given as input;
    \item \texttt{TrackVoltage} produces a file with the value of the voltage. The use of a trace source connected to the variable indicating the remaining voltage may instead have incorrect behavior because of the small difference between consecutive updates, which may not be detected by ns-3 native implementation.
\end{itemize}

\subsection{Variable Energy Harvester}
The class \texttt{VariableEnergyHarvester} extends the \texttt{EnergyHarvester} class provided by ns-3, taking harvested power values from a .csv file given as input. The harvested power is periodically updated, and the energy source object(s) (e.g., the  capacitor) connected to the harvester are updated accordingly. 

The implementation of the function reading the .csv file is specific to the file we considered as input to our scripts. The code can be easily modified to consider data saved in a different format, since it only needs a pair (timestamp, $P_{harvester}$). Note also that, when running simulations, the length of the trace provided to the energy harvester has to be taken into account.

\subsection{Integration with an existing module: \texttt{lorawan}}
To validate the capacitor model, we employ the \texttt{lorawan} ns-3 module~\cite{capuzzo2018confirmed, magrin2017performance}. In particular, we extended the class \texttt{Lo\-ra\-Ra\-dio\-E\-ner\-gy\-Mo\-del}, a child class of the \texttt{De\-vice\-E\-ner\-gy\-Mo\-del}, to work with the capacitor's implementation, and added some states and variables to improve its compliance to real devices. Furthermore, the module's classes representing the \gls{mac} and \gls{phy} layers of the devices have been modified to update and verify the energy level before switching to a new state. The on/off behavior is implemented as follows: if the stored voltage is below $V_{th\_low}$, the capacitor enters in the ``depleted'' state, the \texttt{Lo\-ra\-Ra\-dio\-E\-ner\-gy\-Mo\-del} is notified, and the device enters into the Off state. Conversely, when enough energy is harvested, and the voltage is above $V_{th\_high}$, the capacitor switches from the ``depleted'' to the ``recharged'' state, trigging some events from the \texttt{Lo\-ra\-Ra\-dio\-E\-ner\-gy\-Mo\-del} class (usually, the switch to the Sleep state) and enabling again packet transmission.

Furthermore, we also implemented the behavior of a smarter device, which is able to predict the energy cost of a packet transmission: if the predicted voltage after the operation is below $V_{th\_low}$, the transmission at the \gls{mac} layer is not performed since, being incomplete, it would be unsuccessful.

\label{sub:code} 

\section{Validation}
\label{sec:validation}
In this section, we consider the LoRaWAN technology to validate the \texttt{CapacitorEnergySource} class presented in Section~\ref{sec:implementation}. In the following, we first introduce the LoRaWAN technology and the states that characterize the device operations. Then, we show preliminary results to validate the implementation, which are then expanded to analyze the mutual relations between capacitor's properties and the configuration of the technology, and how they impact on the success of the communication.

\subsection{LoRaWAN}
LoRaWAN~\cite{lorawan} leverages the proprietary LoRa modulation, based on the \gls{css} technique.
The robustness of the modulation can be adjusted by tuning the \gls{sf}
parameter, which takes integer values from 7 to 12. The \gls{sf} is directly
connected to the data rate: higher \glspl{sf} allow for more robust transmitted signals and longer coverage ranges, but at the price of a lower
data rate and, thus, longer transmission times.

LoRaWAN networks have a star-of-stars topology with three
kinds of devices:

\glsreset{ns} \glsreset{ed} \glsreset{gw}
\begin{itemize}
\item \glspl{ed} are peripheral nodes, usually sensors or actuators, that
  communicate using the LoRa modulation;
\item \glspl{gw} are relay nodes that collect messages coming from the
  \glspl{ed} through the LoRa interface, and forward them to the
  \acrlong{ns} using a reliable IP connection, and vice-versa;
\item the \gls{ns} acts as a central network controller that
  manages the communication with the \glspl{ed} through the \glspl{gw}.
\end{itemize}


The LoRaWAN specifications define three classes of \glspl{ed}, which differ
in terms of energy saving capabilities and reception availability. In this work, we focus on Class A devices, which
stay in sleep mode most of the time in order to minimize the energy
consumption, transmit a packet whenever required by the application
layer, and open at most two reception windows after each transmission. If a \gls{dl} message is received in the \gls{rx1}, the \gls{rx2} is not opened.

Moreover, the specifications define two types of messages: confirmed and
unconfirmed. For the first case, an \gls{ack} packet is expected by the
\glspl{ed}, in either of the two reception windows opened after the
transmission. \glspl{ed} have $m$ transmission opportunities: if the
\gls{ack} is not received, the device can re-transmit
the packet up to $m-1$ times. Unconfirmed messages, instead, do not require
any \gls{ack}.

LoRaWAN operates in the unlicensed \gls{ism} frequency bands. The relevant
regulations define frequency bands, power and \gls{dc} restrictions to be
applied. 
In the European region, LoRaWAN~\cite{lorawan} defines the use of three 125~kHz wide channels,
centered at 868.1, 868.3, 868.5~MHz, which are shared between \gls{ul} and
\gls{dl} transmissions, and must collectively respect a 1\% \gls{dc}
constraint~\cite{rec7003e}. As per the specification~\cite{lorawan}, a
fourth 125~kHz channel centered at 869.525~MHz is used for \gls{dl}
communication only, and is subject to a \gls{dc} constraint of
10\%~\cite{rec7003e}. Once fixed the channel bandwidth, there is a one-to-one correspondence between \glspl{sf} and data rates: \gls{sf}~7 corresponds to \gls{dr}~5, \gls{sf}~8 to \gls{dr}~4, and so on, till the most robust \gls{sf}, which corresponds to \gls{dr}~0.


For Class A devices, the standard requires the \gls{rx1} to be opened in the
same frequency channel and with the same \gls{sf} used for the \gls{ul}
communication. The \gls{rx2}, instead, is always opened on the dedicated
869.525~MHz channel and with \gls{sf} 12, in order to maximize the robustness of the communication.

In the following discussion, we assume only \gls{ul} traffic. Furthermore, to measure the quality of the communication, we will indicate as \emph{UL cycle} the interval between the beginning of the \gls{ed}'s packet transmission till the moment when it is successfully delivered to the \gls{gw}, and \emph{UL + DL cycle} the interval from the moment when the \gls{ed} starts the \gls{ul} packet transmission till the successful reception of the corresponding \gls{ack}.

\begin{table}[tb]
    \centering
    \begin{tabular}{llll}
    \toprule
    State & MCU & Radio current & Total current \\
    \midrule
    Off &  Standby & 0 & 5.5 $\mu$A \\
    Turn On & Active & - & 15 mA \\
    Sleep & Active & 1 $\mu$A & 5.6 $\mu$A \\
    Tx & Active & 28 mA & 28.011 mA \\
    Idle & Standby & 1.5 $\mu$A & 7 $\mu$A\\
    Standby & Standby & 10.5 mA & 10.5055 mA\\
    Rx & Active & 11 mA & 11.011 mA \\
    \bottomrule
    \end{tabular}
    \caption{Current consumption in the different states. The current consumption due to \gls{mcu} is 11$\mu$A in active state, 5.5 $\mu$A in standby state.}
    \label{tab:current}
    \vspace{-0.5cm}
\end{table}
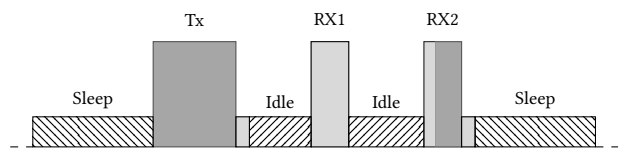
\begin{figure}[t]
  \centering
  \begin{subfigure}{\columnwidth}
    {
      \usetikzlibrary{patterns}


\begin{tikzpicture}

  \def\bottomy{0};
  \def\dashedsidelength{0.3};
  \def\blockheightshort{0.4};
  \def\blockheighttall{1.4};
  \def\blocksleepwidth{1.6};
  \def\blocktxwidth{1.1};
  \def\blockstandbywidth{0.18};
  \def\blockidlewidth{1.};
  \def\blockrxwinwidth{0.5};
  
  \def\ybottom{1};
  \def\textbottom{0};

  \draw [dashed] (0,\ybottom) -- (\dashedsidelength,\ybottom);
  \draw [pattern=north west lines](\dashedsidelength,\ybottom) rectangle node[above, yshift=0.2cm]{\footnotesize{Sleep}} ({\blocksleepwidth + \dashedsidelength},{\ybottom + \blockheightshort});
  
  \draw [fill=gray, opacity=0.7](\dashedsidelength+\blocksleepwidth,\ybottom) rectangle node[text=black, opacity=1, above, yshift=0.8cm]{\footnotesize{Tx}} ({\blocktxwidth + \blocksleepwidth + \dashedsidelength},{\ybottom + \blockheighttall});
  
  \draw (\dashedsidelength+\blocksleepwidth + \blocktxwidth,\ybottom) rectangle node[above, yshift=0.2cm]{\footnotesize{}} ({\blockstandbywidth + \blocktxwidth + \blocksleepwidth + \dashedsidelength},{\ybottom + \blockheightshort});
  \draw [fill=gray, opacity= 0.3](\dashedsidelength+\blocksleepwidth + \blocktxwidth,\ybottom) rectangle node[above, yshift=0.2cm]{\footnotesize{}} ({\blockstandbywidth + \blocktxwidth + \blocksleepwidth + \dashedsidelength},{\ybottom + \blockheightshort});
  
  \draw [pattern = north east lines](\blockstandbywidth + \dashedsidelength+\blocksleepwidth + \blocktxwidth,\ybottom) rectangle node[above, yshift=0.2cm]{\footnotesize{Idle}} ({\blockidlewidth + \blocktxwidth + \blocksleepwidth + \dashedsidelength},{\ybottom + \blockheightshort});
  
  \draw (\blockidlewidth + \dashedsidelength+\blocksleepwidth + \blocktxwidth,\ybottom) rectangle node[text=black, opacity=1, above, yshift=0.8cm]{\footnotesize{RX1}} ({\blockrxwinwidth + \blockidlewidth + \blocktxwidth + \blocksleepwidth + \dashedsidelength},{\ybottom + \blockheighttall});
  \draw [fill=gray, opacity= 0.3, text=black](\blockidlewidth + \dashedsidelength+\blocksleepwidth + \blocktxwidth,\ybottom) rectangle node[text=black, opacity=1, above, yshift=0.8cm]{\footnotesize{}} ({\blockrxwinwidth + \blockidlewidth + \blocktxwidth + \blocksleepwidth + \dashedsidelength},{\ybottom + \blockheighttall});
  
  \draw [pattern = north east lines](\blockrxwinwidth + \blockidlewidth + \dashedsidelength+\blocksleepwidth + \blocktxwidth,\ybottom) rectangle node[text=black, above, yshift=0.2cm]{\footnotesize{Idle}} ({\blockrxwinwidth+2*\blockidlewidth + \blocktxwidth + \blocksleepwidth + \dashedsidelength},{\ybottom + \blockheightshort});

  \draw (\blockrxwinwidth + 2*\blockidlewidth + \dashedsidelength+\blocksleepwidth + \blocktxwidth,\ybottom) rectangle node[text=black, opacity=1, above, yshift=0.8cm]{\footnotesize{RX2}} ({2*\blockrxwinwidth + 2*\blockidlewidth + \blocktxwidth + \blocksleepwidth + \dashedsidelength},{\ybottom + \blockheighttall});
  \draw [draw=none, fill=gray, opacity= 0.3](\blockrxwinwidth + 2*\blockidlewidth + \dashedsidelength+\blocksleepwidth + \blocktxwidth,\ybottom) rectangle node[above, yshift=0.8cm]{\footnotesize{}} ({1.3*\blockrxwinwidth + 2*\blockidlewidth + \blocktxwidth + \blocksleepwidth + \dashedsidelength},{\ybottom + \blockheighttall});
  \draw [draw=none, fill=gray, opacity= 0.7](1.3*\blockrxwinwidth + 2*\blockidlewidth + \dashedsidelength+\blocksleepwidth + \blocktxwidth,\ybottom) rectangle node[above, yshift=0.8cm]{\footnotesize{}} ({2*\blockrxwinwidth + 2*\blockidlewidth + \blocktxwidth + \blocksleepwidth + \dashedsidelength},{\ybottom + \blockheighttall});

  \draw (2*\blockrxwinwidth + 2*\blockidlewidth + \dashedsidelength+\blocksleepwidth + \blocktxwidth,\ybottom) rectangle node[text=black, above, yshift=0.2cm]{\footnotesize{}} ({\blockstandbywidth + 2*\blockrxwinwidth + 2*\blockidlewidth + \blocktxwidth + \blocksleepwidth + \dashedsidelength},{\ybottom + \blockheightshort});
  \draw [fill=gray, opacity= 0.3](2*\blockrxwinwidth + 2*\blockidlewidth + \dashedsidelength+\blocksleepwidth + \blocktxwidth,\ybottom) rectangle node[text=black, above, yshift=0.2cm]{\footnotesize{}} ({\blockstandbywidth + 2*\blockrxwinwidth + 2*\blockidlewidth + \blocktxwidth + \blocksleepwidth + \dashedsidelength},{\ybottom + \blockheightshort});

  \draw [pattern=north west lines](2*\blockrxwinwidth + 2*\blockidlewidth + \dashedsidelength+\blocksleepwidth + \blocktxwidth + \blockstandbywidth,\ybottom) rectangle node[above, yshift=0.2cm]{\footnotesize{Sleep}} ({\blockstandbywidth + 2*\blockrxwinwidth + 2*\blockidlewidth + \blocktxwidth + 2*\blocksleepwidth + \dashedsidelength},{\ybottom + \blockheightshort});
  
  \draw [dashed] ({\blockstandbywidth + 2*\blockrxwinwidth + 2*\blockidlewidth + \blocktxwidth + 2*\blocksleepwidth + \dashedsidelength},\ybottom) -- ({\blockstandbywidth + 2*\blockrxwinwidth + 2*\blockidlewidth + \blocktxwidth + 2*\blocksleepwidth + 2*\dashedsidelength},\ybottom);

\end{tikzpicture}
    }
  \end{subfigure}
  \caption{Example of \gls{ed}'s state transitions. In this case, a \gls{dl} packet is received in \gls{rx2}, after a short time spent in standby mode.}
  \label{fig:cycle}
\end{figure}
\begin{figure*}[t]
  \centering
   \begin{subfigure}{0.33\textwidth}
    {    
        \includegraphics[width=\linewidth]{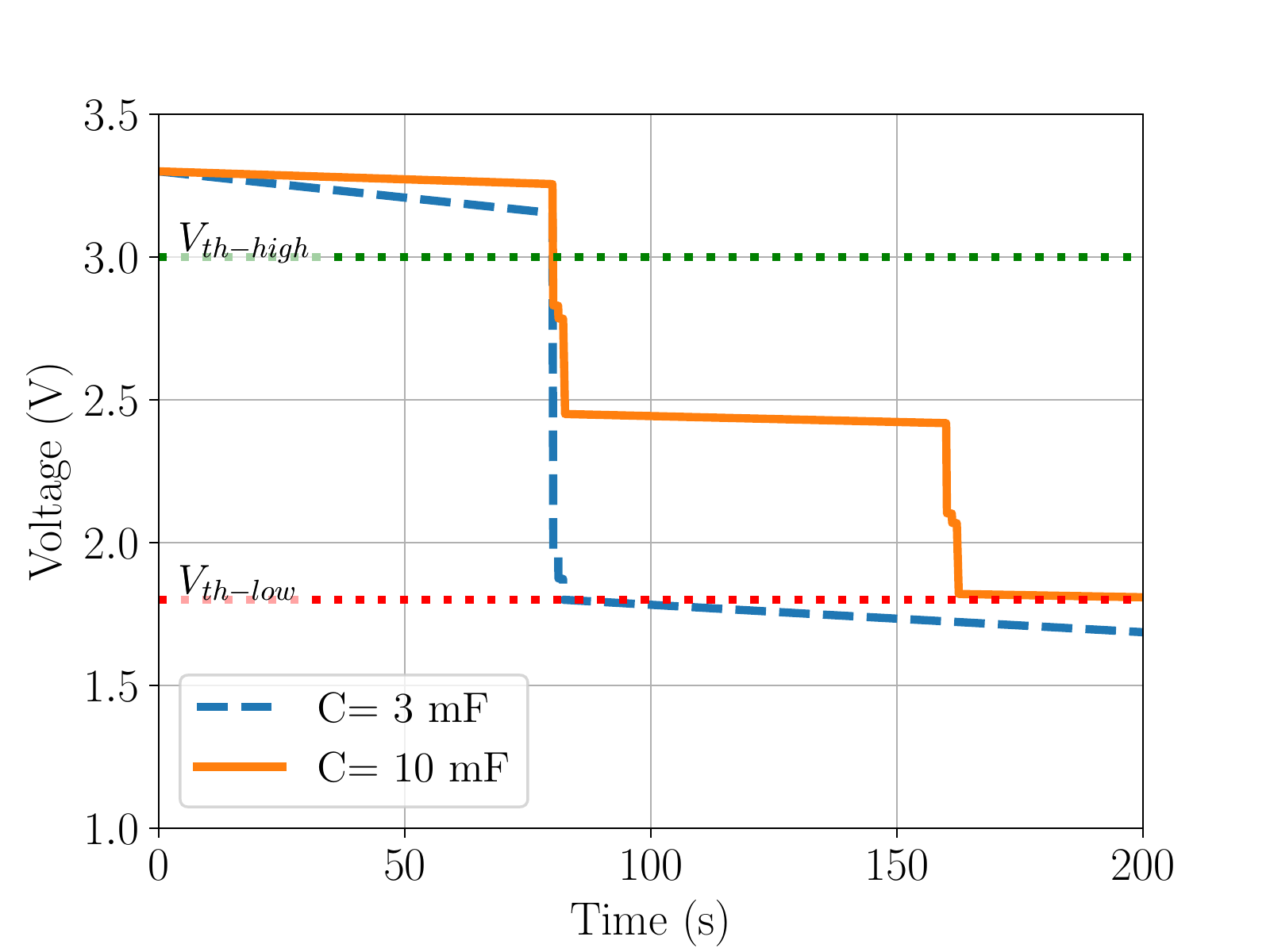} 
        \caption{$P_{harvester}$ = 0 W.}
        \label{fig:eh0}
    }
    \end{subfigure}
    \hfill
    \begin{subfigure}{0.33\textwidth}
    {
        \includegraphics[width=\linewidth]{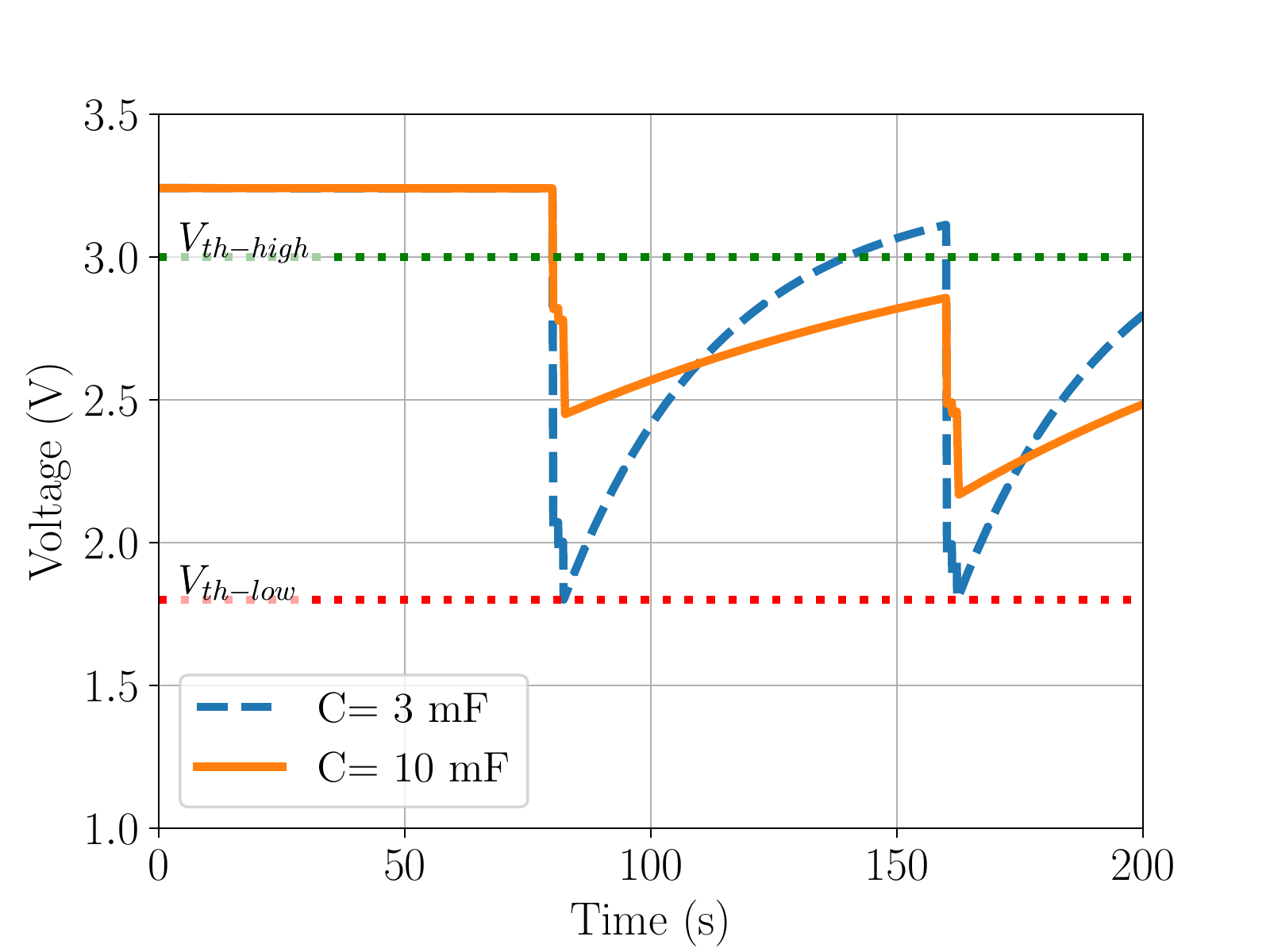} 
        \caption{$P_{harvester}$ = 0.001 W.}    
        \label{fig:eh0001}
    }
    \end{subfigure}
    \hfill
    \begin{subfigure}{0.33\textwidth}
    {
        \includegraphics[width=\linewidth]{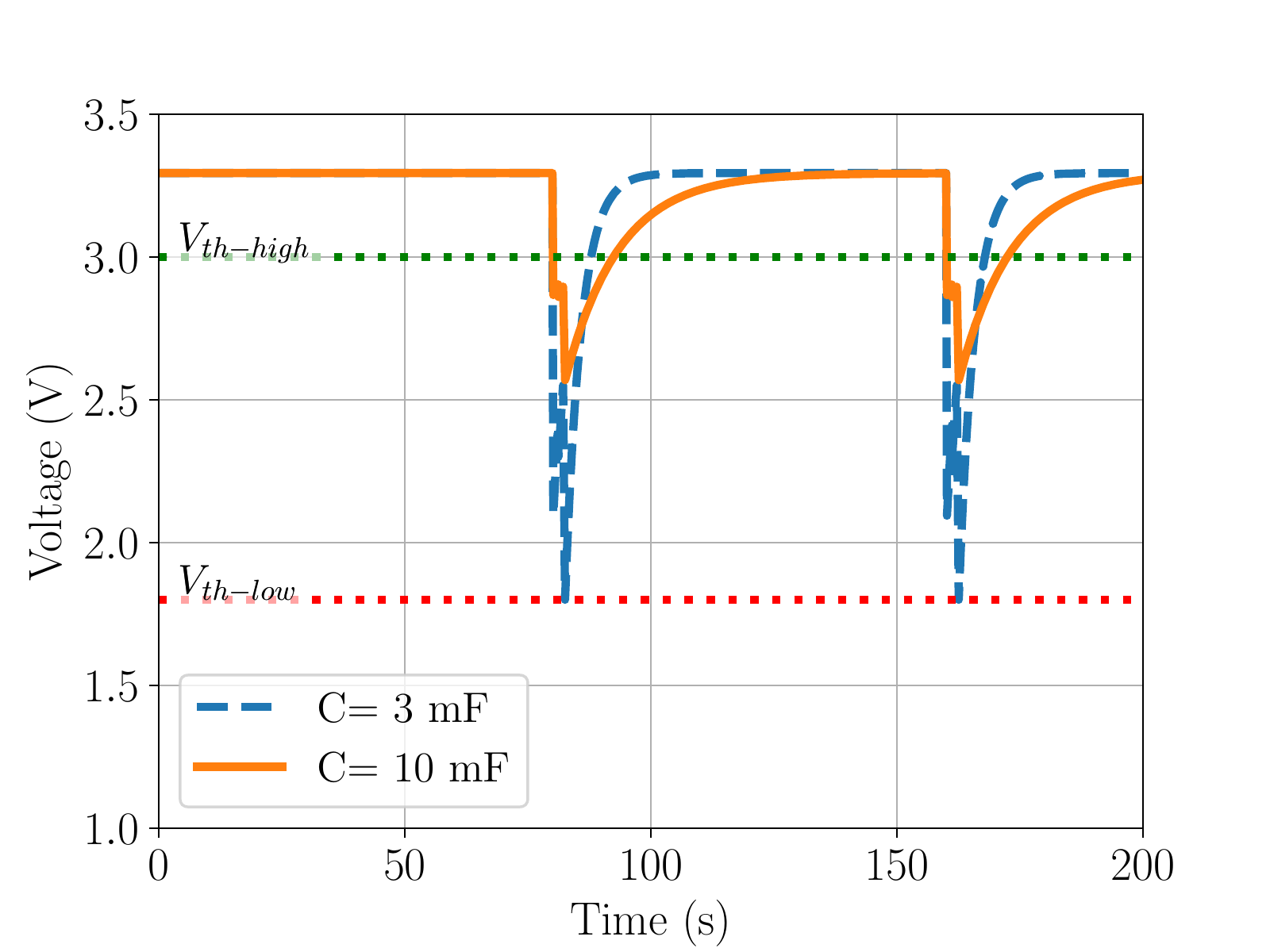} 
        \caption{$P_{harvester}$ = 0.01 W.}    
        \label{fig:eh001}
    }
    \end{subfigure}
  \caption{Voltage for different values of $P_{harvester}$ and capacitor's size.}
  \label{fig:eh}
\end{figure*}

\subsubsection{LoRaWAN device's states}

As discussed in Section~\ref{sec:implementation}, the different states the device goes through are important to determine the energy consumption of the device. According to the LoRaWAN protocol described above, the following states are identified. The respective current consumption is reported in Table~\ref{tab:current}, considering the load composed by the \gls{mcu} and radio units.
\begin{itemize}
    \item \emph{Off}(*): the \gls{ed}'s radio is switched off, and the \gls{mcu} is in standby, maintaining only the clock synchronization;
    \item \emph{TurnOn}(*): the device wakes up from the Off state, with a certain energy expenditure. In our implementation we consider a current consumption of 15~mA and a state duration of 300~ms, but the values can be tuned according to specific devices considered;
    \item \emph{Sleep}: the radio is in sleep state, saving power, without performing any activity, and the \gls{mcu} is in standby mode;
    \item \emph{Tx}: the device is transmitting data;
    \item \emph{Idle}: ``waiting'' period before the opening of the receive window;
    \item \emph{Standby}: listening to idle channel when the receive windows are open. Also, the standard defines the \gls{ed} to switch to Standby (for a very short time) after transmission and reception operations;
    \item \emph{Rx}: the device is receiving data.
\end{itemize}

The Off and Turn On states, marked with (*), are not part of the standard, but are present in real devices. In Fig.~\ref{fig:cycle} a diagram depicts the operational states of the device. In this case, the device wakes up from the Sleep state to transmit data requiring an \gls{ack}, which is successfully received in \gls{rx2}. The light-grey-colored regions represent the Standby phase.
Note that the device will be able to complete a \gls{ul} cycle if the voltage level is enough to complete the transmission procedure (no impairments from the channel/network are considered), while a \gls{ul} + \gls{dl} cycle is successfully executed if the energy stored in the device is enough to complete all the operations from transmission to reception, which can happen in either \gls{rx1} or \gls{rx2}, including also the intermediate Idle and Standby states.

\subsection{Results}
To validate our ns-3 implementation, we consider a simple LoRaWAN network composed of a \gls{ns}, a \gls{gw} and a single \gls{ed} provided with a capacitor with variable size. Furthermore, we consider $V_{th\_low}$=1.8~V and $V_{th\_high}$=3~V, and different values for $P_{harvester}$, from 0~W to 0.01~W, which are constant in time. The device periodically generates packets with a payload length of 10~bytes, and can use either unconfirmed or confirmed messages, with $m$=1. The smart option preventing a packet transmission if the energy cost is not supported by the device is used. The LoRa settings are as considered in~\cite{delgado2020battery}.

Fig.~\ref{fig:timevoltage} shows the capacitor's voltage together with the states of the device.\footnote{Markers signaling when entering in Standby and Idle states are not plotted for clarity.} The initial volage of the capacitor is 3.3~V, which is almost constant for the first part of the simulation, when the device is in Sleep mode. At $t=$80~s, the \gls{ed} performs the first transmission, entering the Tx state, which causes a first drop in the capacitor's voltage, bringing it to 2.44~V. The traffic type is unconfirmed, therefore, no \gls{dl} transmission is expected. Nonetheless, as dictated by the standard, the two reception windows are opened, which cause the voltage drop around time 81~s. Note that, as it can be seen also in Fig.~\ref{fig:eh0}, the energy drop during \gls{rx1} is smaller than that experienced during \gls{rx2}, whose duration is $2^6=64$ times longer than that of \gls{rx1}. In the simulated scenario, the harvesting rate is EH=0.001~W, which allows the capacitor to recharge during the sleeping period from 81~s to 160~s, reaching almost 3~V. Note that this is not visible during the initial sleeping period, because the voltage level is very close to the maximum voltage supported by the device. At 160~s a second transmission occurs, followed by the opening of the two reception windows. In this case, the voltage at the beginning of the cycle was lower than in the previous case, and the long duration of \gls{rx2} makes the voltage drops below $V_{th\_low}$, so that the \gls{ed} enters the Off state. Then, a recharging phase follows, and when $v(t, s)$ reached $V_{th\_high}=$3~V, the \gls{ed} starts the TurnOn phase, entering into the Sleep state. This enables it to successfully perform the next transmission, at time $t$=240~s.

Fig.~\ref{fig:eh} shows the voltage level for different capacitor sizes and $P_{harvester}$ values, which have a  determinant impact on the communication capabilities. As a benchmark, in Fig.~\ref{fig:eh0} we report the case with no harvesting. from this, we can appreciate the impact of different capacitors' sizes: a smaller capacitor discharges much faster than a bigger one, and can rapidly make the device switch off. However, when using smaller capacitors, also the recharging phases are faster, as it can be observed in Figs.~\ref{fig:eh0001} and \ref{fig:eh001}: this behavior may cause the device to swap between On and Off states, preventing proper communication. Conversely, a larger capacitance will charge and discharge more slowly, allowing better communication performance (in terms of successful transmissions) also in the case of lower harvesting rates, since it will reduce the number of times the node enters in Off state and, consequently, the energy cost for taking it back to the active state. The downside is that, whenever the capacitor voltage drops below the lower threshold, it will take a longer time to accumulate enough energy to pass again the high threshold and bring the device back to an operational state.

\begin{figure}[t]
  \centering
   \begin{subfigure}{0.45\textwidth}
    {
        \includegraphics[width=\linewidth]{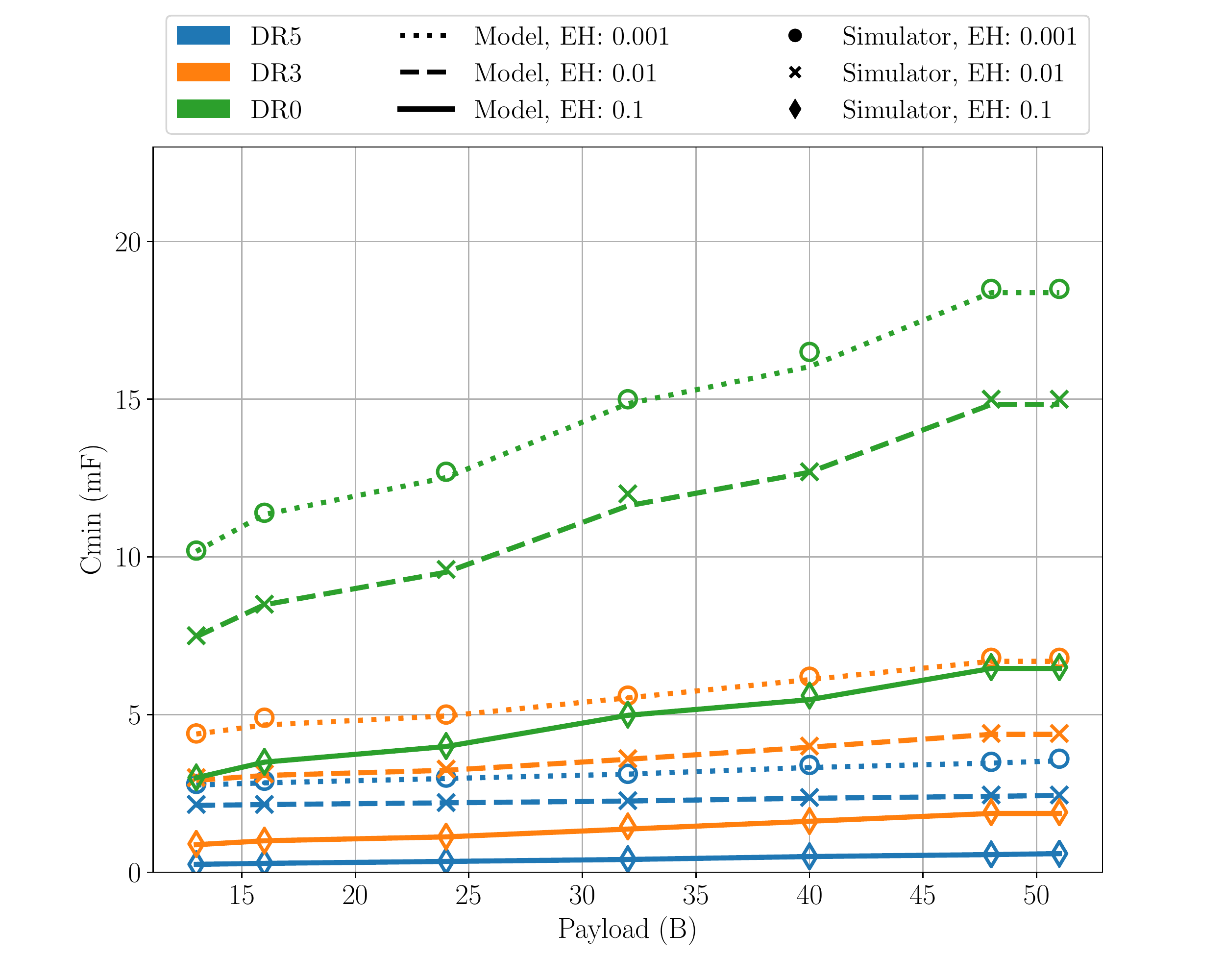} 
        \caption{UL cycle.}    
    }
    \end{subfigure}
    \begin{subfigure}{0.45\textwidth}
    {
        \includegraphics[width=\linewidth]{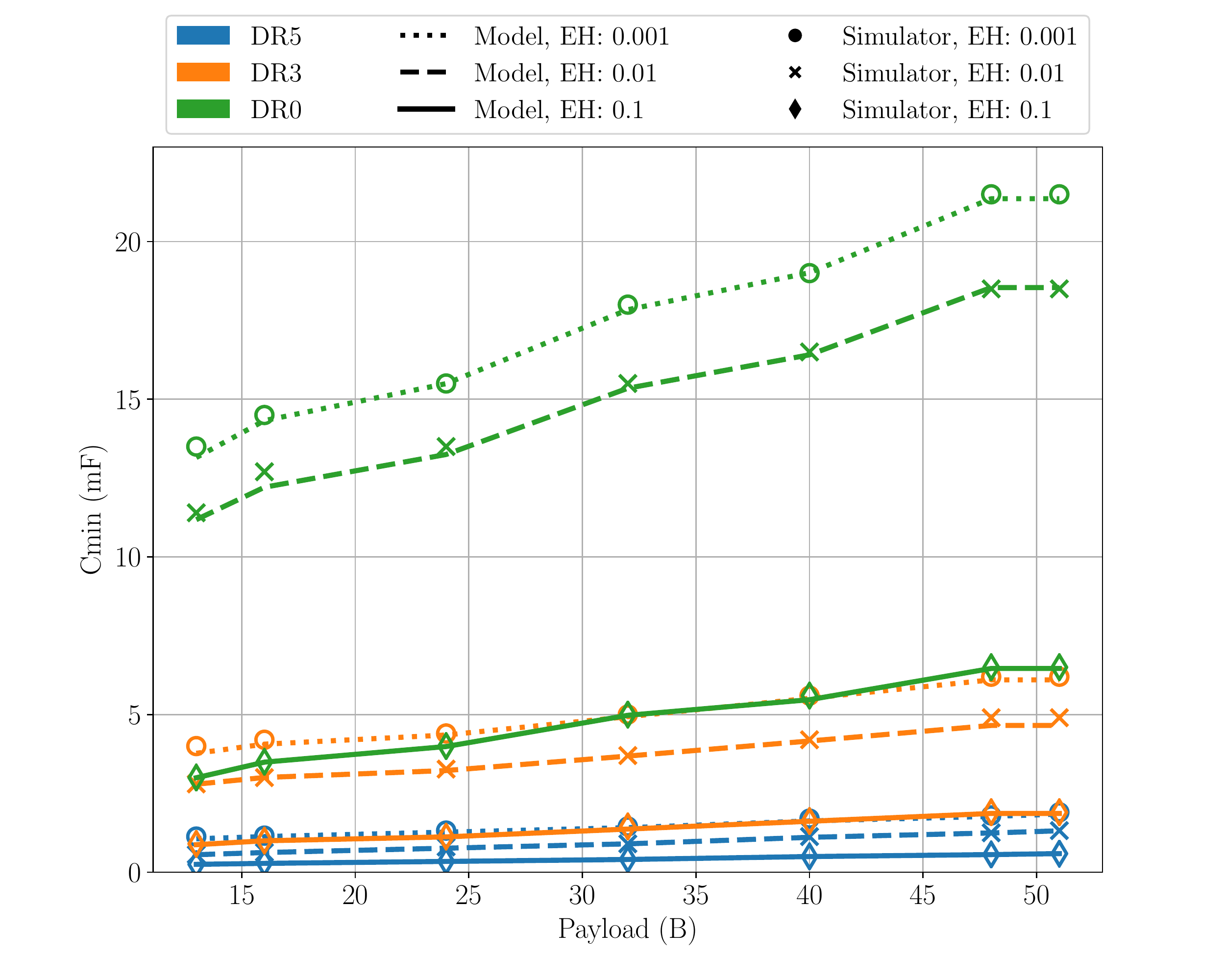} 
        \caption{UL and DL cycle.}    
    }
    \end{subfigure}
  \caption{Minimum capacity to complete a cycle. Comparison with mathematical results, computed as in~\cite{delgado2020battery}.}
  \label{fig:minc}
\end{figure}

The minimum capacity size that makes it possible to complete a \gls{ul} (resp. \gls{ul} + \gls{dl}) cycle for different \gls{ul} packet sizes, harvesting rates and \glspl{dr} is presented in Fig.~\ref{fig:minc}, and compared with mathematical results obtained from the model proposed in~\cite{delgado2020battery}, which are represented with diamond markers and gray line. The \gls{dl} packet size is fixed to 39~bytes (at APP layer). Also, the initial capacitor voltage is computed taking into account the current in the Off state. Since there is a single \gls{ed} in the network, the \gls{gw} can always use \gls{rx1}, sparing the device the energy consumption due to additional states. 
In both plots, we can observe that the minimum required capacitance increases for bigger payloads of the \gls{ul} packet, as expected. Moreover, the lower the \gls{dr}, the larger the required capacitance, because of the longer transmission time. For lower harvesting rates, a larger capacitor should be employed to successfully complete a cycle, as discussed previously. Similar trends can be observed for the minimum capacitance needed to accomplish a \gls{ul} + \gls{dl} cycle: in this case, the values are higher than for the \gls{ul} cycle only, since additional actions must be performed by the device, including the reception of a \gls{dl} packet. From these plots we can finally observe that there is a strong agreement between model and simulation results, with the small discrepancy between the two only due to the quantized step used in the simulator.
%

\begin{figure*}[t]
  \centering
   \begin{subfigure}{0.42\textwidth}
    {
        \includegraphics[width=\linewidth]{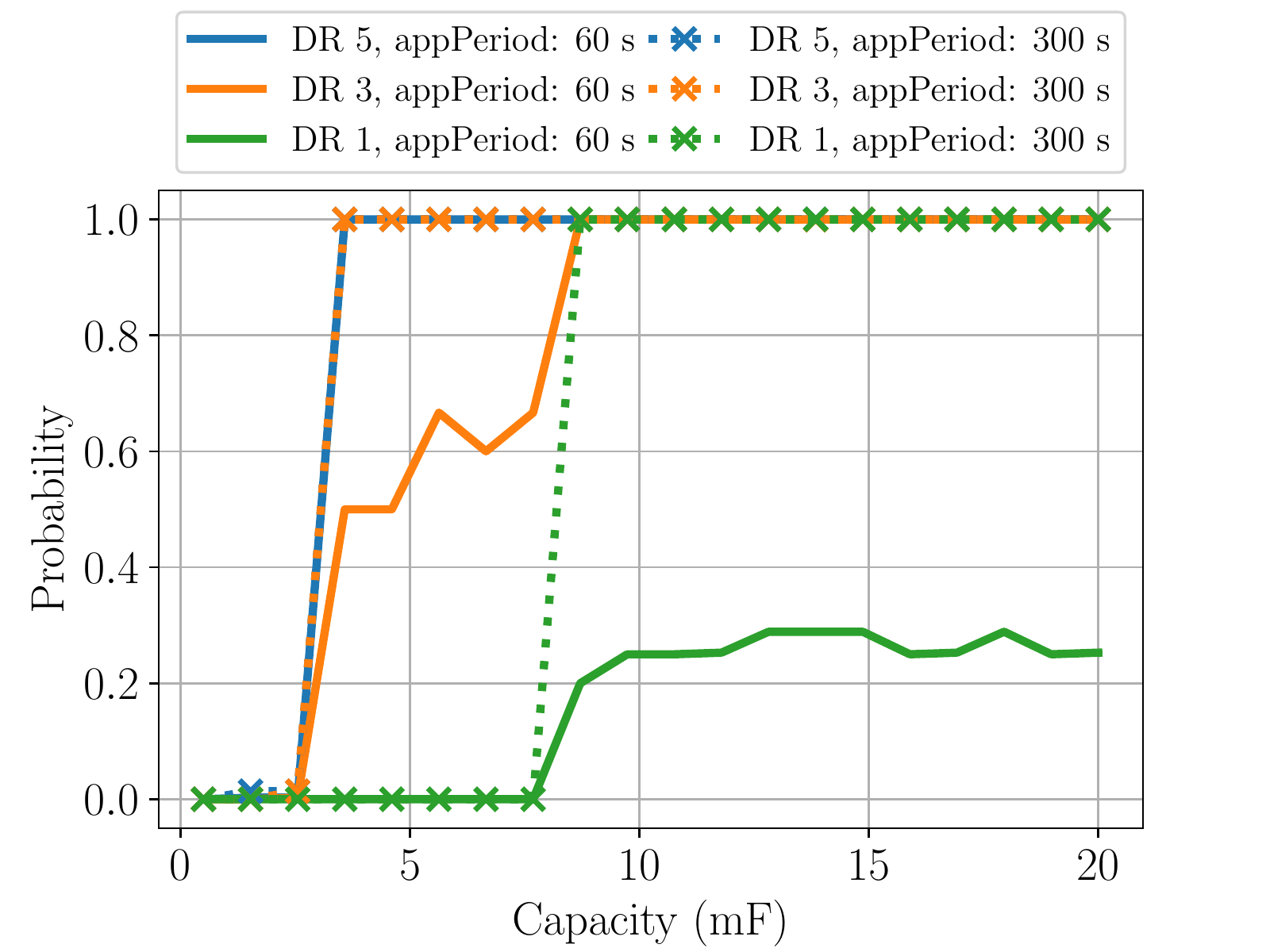} 
        \caption{$P_{harvester}$ = 0.001 mW.}    
        \label{fig:psuccunconf0001}
    }
    \end{subfigure}
    \begin{subfigure}{0.42\textwidth}
    {
        \includegraphics[width=\linewidth]{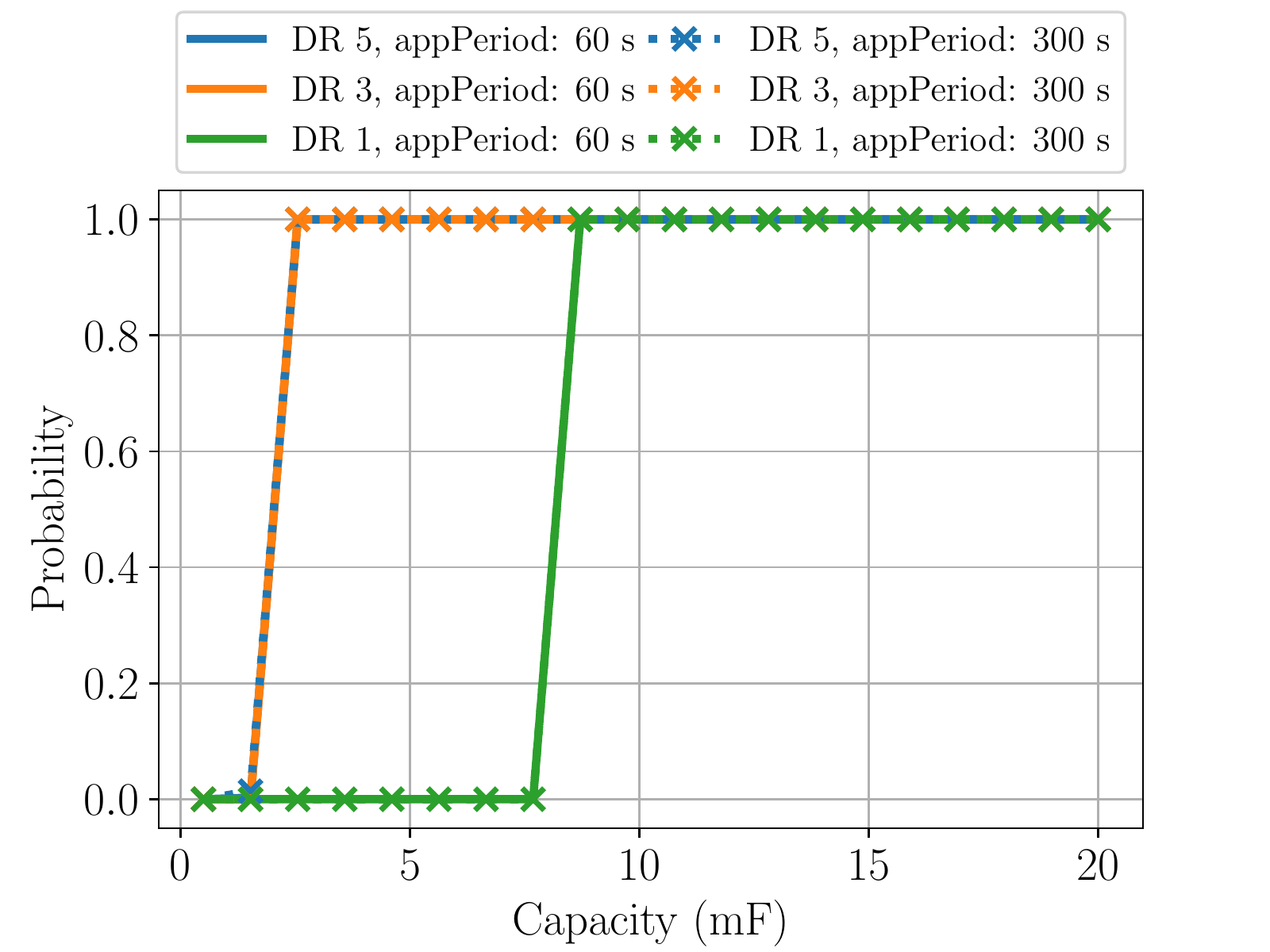} 
        \caption{$P_{harvester}$ = 0.01 mW.} 
        \label{fig:psuccunconf001}
    }
    \end{subfigure}
    \caption{Success probability for unconfirmed traffic.}
  \label{fig:psuccunc}
\end{figure*}

\begin{figure*}[t]
  \centering
   \begin{subfigure}{0.42\textwidth}
    {
        \includegraphics[width=\linewidth]{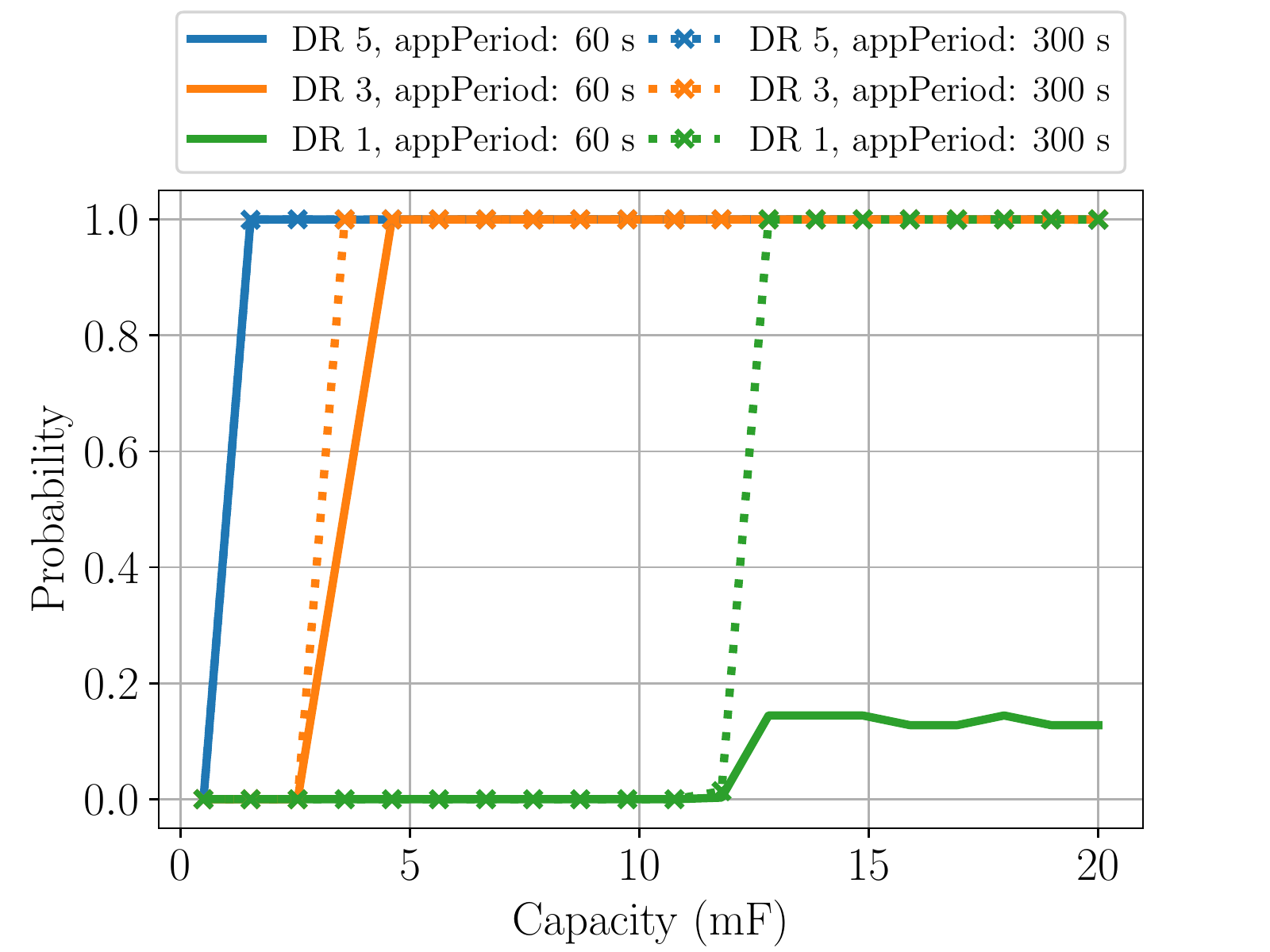} 
        \caption{$P_{harvester}$ = 0.001 mW.}   
        \label{fig:psuccconf00001}
    }
    \end{subfigure}
    \begin{subfigure}{0.42\textwidth}
    {
        \includegraphics[width=\linewidth]{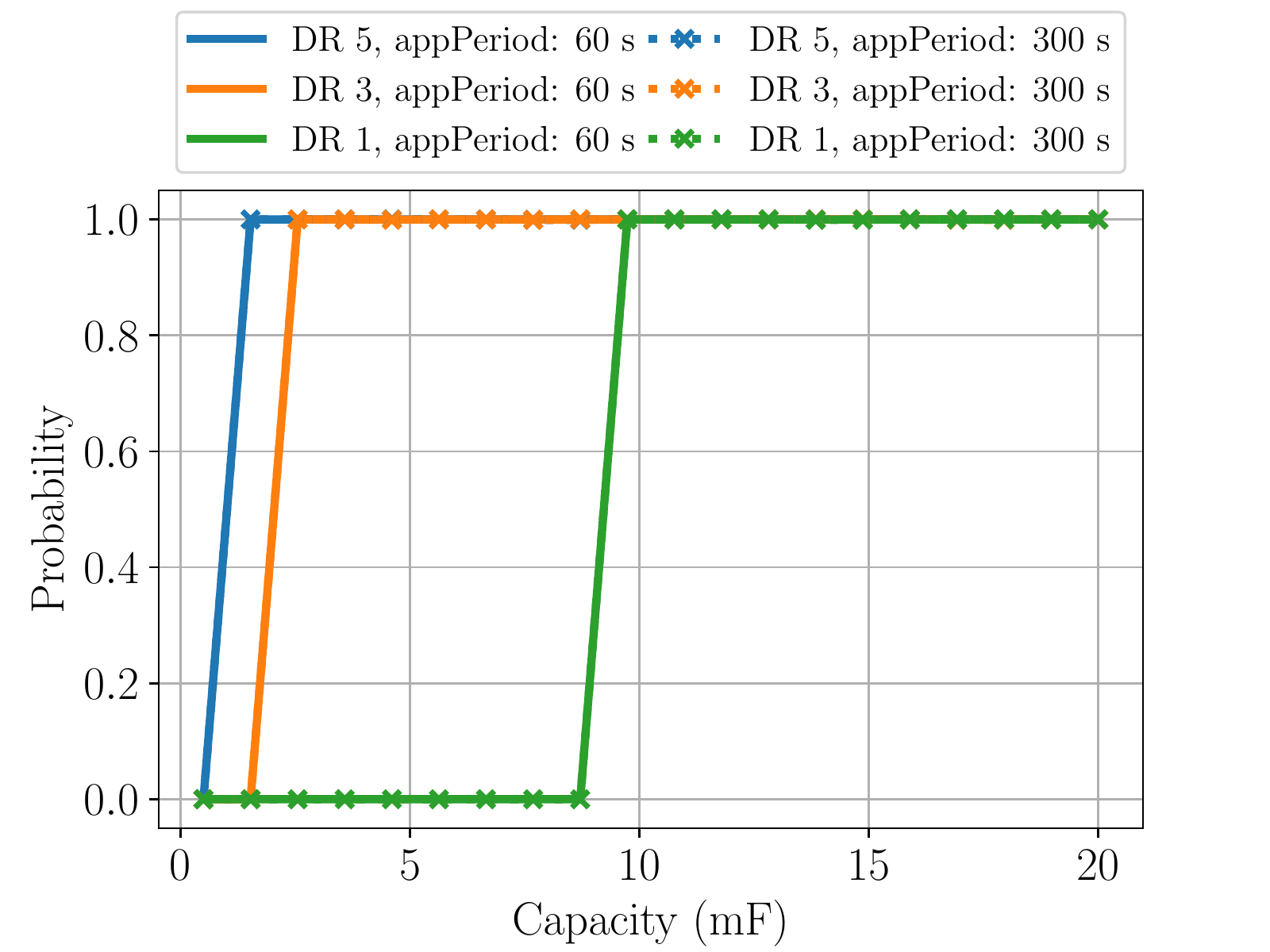} 
        \caption{$P_{harvester}$ = 0.01 mW.}    
    }
    \end{subfigure}
  \caption{Success probability for confirmed traffic.}
  \label{fig:psuccconf}
\end{figure*}

A final batch of simulations was run to reproduce a realistic scenario where an \gls{iot} application periodically generates packets of fixed size. In particular, we evaluated the success probability (in terms of delivered packets) when varying the capacitor size, for an application sending confirmed/unconfirmed traffic for 6~hours, with different packet generation periods. Note that the success probability is computed as the ratio between the number of delivered packets and the total number of packets generated at the application level.
Figs.~\ref{fig:psuccunc},~\ref{fig:psuccconf} show the results for different values of the packet generation period and harvesting rate.
Also in this case, we can observe a strong dependence on the \gls{dr}: while lower \gls{dr} values improve the transmission robustness to possible channel impairments, they are more costly in terms of energy, strongly affecting the number of packets that are successfully received by the \gls{gw}. This can be mitigated by storing more energy, as happens for bigger capacitors charged with higher harvested power (Fig.~\ref{fig:psuccunconf001}). Instead, using higher \gls{dr} values require smaller capacitors, in the order of a few mF. Furthermore, transmitting packets more sporadically leaves enough time to recharge the capacitor, obtaining a higher success probability for a given capacitor's size. For example, increasing the interval between consecutive packet transmissions from 60~s to 300~s makes it possible to halve the minimum capacitance when using \gls{dr} 3 and $P_{harvester}=0.001$~mW (Fig.~\ref{fig:psuccunconf0001}).
Fig.~\ref{fig:psuccconf} reports similar results for the probability of also receiving the \gls{ack}: in this case, similar considerations on the relation between minimum capacity, \gls{dr} and $P_{harvester}$ values can be drawn. However, it is interesting to note that, for low values of $P_{harvester}$ (Fig.~\ref{fig:psuccconf00001}), the capacitor's size that maximizes the success probability is lower in the case of confirmed traffic than when using unconfirmed traffic, despite the reception og \gls{dl} packets occurs. This confirms that using an \gls{ack} (with no payload) to prevent the opening of \gls{rx2} brings some benefit on the \gls{ed}'s energy consumption and communication performance, specially when the harvested power is low. This aspect could be taken into account for a proper network configuration that targets energy efficiency: using shorter \gls{rx2} by employing higher \glspl{dr}, or even preventing their use, could have a significant effect on the device's energy performance.

\section{Conclusions}
\label{sec:conclusions}
In this paper, we propose an implementation of a (super-)capacitor that considers the presence of a generic harvesting source to power battery-less devices. After presenting the considered model, we describe the code implementation and validate it by simulating a simple LoRaWAN network with a single device. The correctness of the implementation was proved by the comparison of some outcomes with those obtained in previous theoretical works. From the results obtained, it is apparent the impact of the harvested power and capacitor size on the communication performance, as well as the effect of the technology settings on the energy requirements. 
The possibility of including the proposed model in the simulation ecosystem makes it possible to further study how the harvesting approach influences the communication performance and, at the same time, it allows for the performance evaluation of a complete \gls{iot} network where many energy-constrained devices compete for the same resources. 

The ns-3 code of the capacitor implementation is available at~\cite{ns3capacitor}.

\begin{acks}
A special thanks goes to Davide Magrin for his useful suggestions. \\
Part of this research was funded MIUR (Italian Ministry for Education and
Research) under the initiative "Departments of Excellence" (Law 232/2016), and by the Flemish FWO SBO S001521N IoBaLeT
(Sustainable Internet of Battery-Less Things) and the University of Antwerp IOF funded COMBAT (Time-Sensitive Computing on Battery-Less IoT Devices) projects.
\end{acks}

\bibliographystyle{ACM-Reference-Format.bst}
\bibliography{biblio.bib}

\end{document}